\documentclass[aps,prl,twocolumn,english,superscriptaddress,showpacs,floatfix]{revtex4}

\usepackage{amsmath,bbm}
\usepackage{psfig,graphics,graphicx}
\usepackage{babel}

\begin{document}

\title{From ballistic to Brownian vortex motion in complex oscillatory media}

\author{J\"orn Davidsen}
\affiliation{Max-Planck-Institut f\"ur Physik Komplexer Systeme,
N\"othnitzer Strasse 38, 01187 Dresden, Germany}
\author{Ronaldo Erichsen}
\affiliation{Universidade Estadual do Rio Grande do Sul, Porto
Alegre, Brazil}
\author{Raymond Kapral}
\affiliation{Max-Planck-Institut
f\"ur Physik Komplexer Systeme,
N\"othnitzer Strasse 38, 01187 Dresden, Germany}
\affiliation{Chemical Physics
Theory Group, Department of Chemistry, University of Toronto,
Toronto, ON M5S 3H6, Canada}
\author{Hugues Chat\'e}
\affiliation{CEA --- Service
de Physique de l'Etat Condens\'e, Centre d'Etudes de Saclay, 91191
Gif-sur-Yvette, France}

\date{\today}

\begin{abstract}
We show that the breaking of the rotation symmetry of spiral waves
in two-dimensional complex (period-doubled or chaotic) oscillatory
media by synchronization defect lines (SDL) is accompanied by an
intrinsic drift of the pattern. Single vortex motion changes from
ballistic flights at a well-defined angle from the SDL to
Brownian-like diffusion when the turbulent character of the medium
increases. It gives rise, in non-turbulent multi-spiral regimes,
to a novel ``vortex liquid''.
\end{abstract}

\pacs{82.40.Ck,45.70.Qj,89.75.Kd,82.40.Bj}

\maketitle

Chemical waves in reaction-diffusion systems with local excitable
or simple oscillatory dynamics have been investigated extensively
both theoretically and experimentally because of their relevance
for a variety of physical, chemical, and biological processes
\cite{kapral,mikhailov}. In (quasi) two-dimensional situations,
spiral wave patterns are especially prevalent and important. They
determine the characteristics of processes such as surface
catalytic oxidation reactions \cite{imbihl95}, contraction of the
heart muscle \cite{winfree87}, and various signalling mechanisms
in biological systems \cite{goldbeter}, to name only a few
examples. The dynamics of the core or ``vortex'' of spiral waves
plays an important role in many of these phenomena: some
mechanisms for spiral breakup arise from core motion (see, for
example, Ref.~\cite{brusch03}) and moving spirals have been
suggested to be responsible for some cardiac arrhythmias
\cite{moveheart}. However,  whereas the meandering instability of
spiral cores in excitable media is well known \cite{barkley95},
studies of vortex motion are still few in the oscillatory case,
being limited to the unbounded acceleration characteristic of the
core instability \cite{aranson02} or to erratic motion induced by
spatiotemporal chaos or external noise \cite{aranson98,alonso01}.

When the local oscillations are not simple but possess a complex
periodic or chaotic character as observed, for example, in
chemically reacting systems \cite{scott}, spiral waves contain
synchronization defect lines (SDL) \cite{goryachev00}
separating domains of different oscillation phases and across which
the phase changes by multiples of $2\pi$. SDL have been observed
in experiments on Belousov-Zhabotinsky reaction.
\cite{yoneyama95,park99} While their origin and classification
have been investigated, little is known about how they influence
the spiral core to which they are connected.

In this Letter, we show that in complex oscillatory media the
emergence of SDL is accompanied by spiral core motion. The SDL
breaks the rotational symmetry of the spiral wave giving rise to a
generic mechanism for core motion that differs from the
instabilities that cause the non-saturating core instability in
simple oscillatory media and the meandering instability in simple
excitable media. We show that spirals in complex oscillatory media
move ballistically in directions that are determined by the SDL
attached to the cores. In the regime of SDL-mediated turbulence
where SDL are spontaneously generated, the core motion is more
complicated leading to randomly-oriented flights of random
duration. With stronger turbulence this dynamics leads to vortex
Brownian motion characterized by a well-defined diffusion
constant. We finally show that multi-spiral configurations in the
locally non-turbulent regime lead to a spatiotemporally chaotic
``vortex liquid''.

\begin{figure}
 \includegraphics[width=\columnwidth]{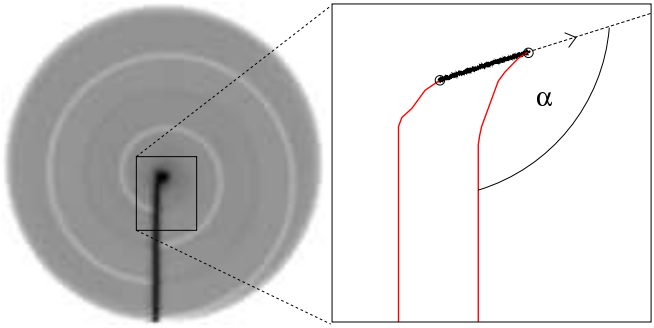}
 \caption{\label{line} (color online) Left: Snapshot of the scalar
 field $\Delta c_z({\bf r},t) = 1/\tau \int^{\tau}_0
 |c_z({\bf r},t+t')-c_z({\bf r},t+t'+\tau)|dt'$ in the P2
 regime for $C=3.5$, $L=256$, $D=0.4$ (kept fixed throughout).
 The period of the P1 oscillation is
 $\tau=5.95$. Black corresponds to $\Delta c_z \approx 0$
 and indicates a SDL originating from the spiral core.
 Right: Magnification of the rectangular region in the left panel
 showing the SDL (red lines) at two times. $\alpha$ is defined
 as the angle between the
 spiral core's trajectory (thick black line) and the attached SDL.
 }
 \end{figure}

Consider the reaction-diffusion system
\[
\partial_t {\bf c}({\bf r},t) = {\bf R}({\bf c}({\bf r},t)) +
D\nabla^2 {\bf c}({\bf r},t),
\]
where ${\bf c}({\bf r},t)$ is a vector of time-dependent
concentrations at point ${\bf r}$ in a two-dimensional domain
of length/diameter $L$ \cite{size}, $D$ is the diffusion
coefficient (taken to be the same for all species) and where the
local kinetics is specified by the vector-valued function ${\bf
R}({\bf c}({\bf r},t))$. As a paradigmatic example of a system
with complex local dynamics we take ${\bf R}({\bf c})$ to be given
by the R\"ossler model \cite{roessler76} with $R_x=-c_y-c_z$,
$R_y=c_x+0.2c_y$, $R_z=c_xc_z-Cc_z+0.2$. For $C \in [2.0,6.0]$,
the medium undergoes period-doubling
bifurcations transforming the local dynamics from simple
oscillatory to period-doubled orbits to chaotic dynamics. The
rotational symmetry of spiral wave patterns is then broken by the
ineluctable appearance of SDL \cite{goryachev00}: In a simple
oscillatory medium, one turn of the spiral wave corresponds to one
period of the local oscillation in phase space (change of $2\pi$
in the phase). When this spiral wave pattern undergoes a period
doubling bifurcation, the period of the local orbits doubles
(change of $4\pi$ in the phase). Continuity in the medium then
forces the emergence of a narrow strip connected to the core, the
SDL, across which the phase jumps by $2 \pi$. Two oscillations are
needed to return the medium to its original state. Along the SDL,
the local dynamics is period-1 (P1) while in the remainder of the
medium it is period-2 (P2). An example of a P1 SDL in a P2 medium
is shown in the left panel of Fig.~\ref{line}. In the period-4
(P4) regime, two types of SDL are possible: lines where the phase
jumps by $2\pi$ (local P1 dynamics) and lines where the phase
jumps by $4\pi$ (local P2 dynamics). More generally, in a medium
with period $2^n$ dynamics $2^{n-1}$ SDL with periodicities $2^k$,
$(k<n)$, may exist, although for the R\"ossler medium only the P4
regime is observed before the local dynamics becomes chaotic. In
these chaotic regimes the P1 and P2 SDL persist. Deeper in the
chaotic regime turbulent states are found where SDL are
spontaneously created and annihilated. \cite{goryachev00}

 \begin{figure}
 \includegraphics[width=\columnwidth]{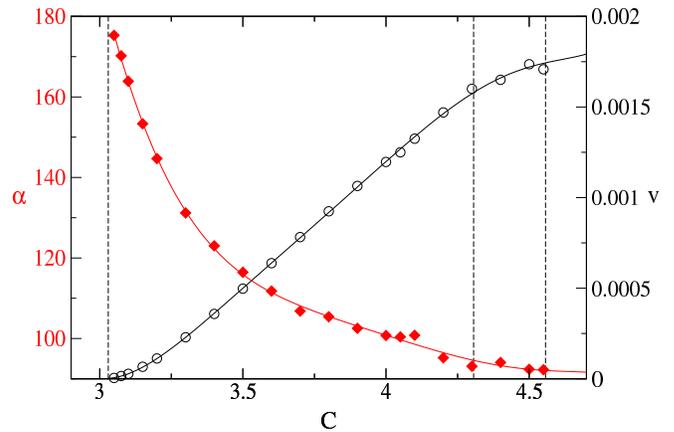}
 \caption{\label{alpha} (color online) Dependence of the angle $\alpha$ (filled diamonds)
 and the core velocity, $v$ (circles) on the R\"ossler parameter $C$.
 The solid lines are guides to the eye.
 From left to right, the vertical lines mark the first period-doubling transition
 to the complex oscillatory regime and the transitions
 to the turbulent regimes (see text for details).
 }
 \end{figure}

First, we consider a single spiral in a disc-shaped domain with
no-flux boundary conditions and focus on the complex oscillatory
regime where no P1 SDL are spontaneously generated ($3.03 < C <
4.557$). Our extensive simulations revealed that  the simplest
possible of the allowed configurations, where a single P1 SDL is
attached to the core, is the unique asymptotic solution, even in
the P4 and chaotic regimes where P2 SDL exist. The presence of the
SDL breaks the rotational symmetry and thus, on general grounds,
one expects the spiral core to move \cite{sandstede}. This is
indeed what is observed, albeit this motion is very slow, taking
typically several thousand oscillations to move by one wavelength.
After transients, the core moves ballistically at constant speed.
In a finite system, it eventually encounters a boundary which
influences its motion. Figure~\ref{alpha} shows that the speed $v$
increases continuously and monotonically from the onset of P2 bulk
oscillations, and analysis of the data indicates a power law
behavior  $v \sim (C-C_2)^\gamma$, where $C_2\approx 3.03$ is the
first period-doubling bifurcation and $\gamma=1.5 \pm 0.02$. The
angle $\alpha$ between the direction of motion and the attached
SDL is constant (Fig.~\ref{line}) and gradually decreases from
$180^\circ$ to $90^\circ$ with increasing $C$ as shown in
Fig.~\ref{alpha}.

In the P4 and higher regimes both P2 and P1 SDL may exist. We
observed that
 both $\alpha$ and $v$ are determined solely by
the attached P1 SDL and are unaffected by any P2 SDL that may be
attached to the core, be it during a transient or in the regime
where P2 SDL are spontaneously and continuously generated ($4.306
\leq C < 4.557$). We conclude that if it exists at all, the effect
of an attached P2 SDL is much weaker than that of the dominant P1
line, and essentially undetectable.

 \begin{figure}
 \includegraphics[width=\columnwidth]{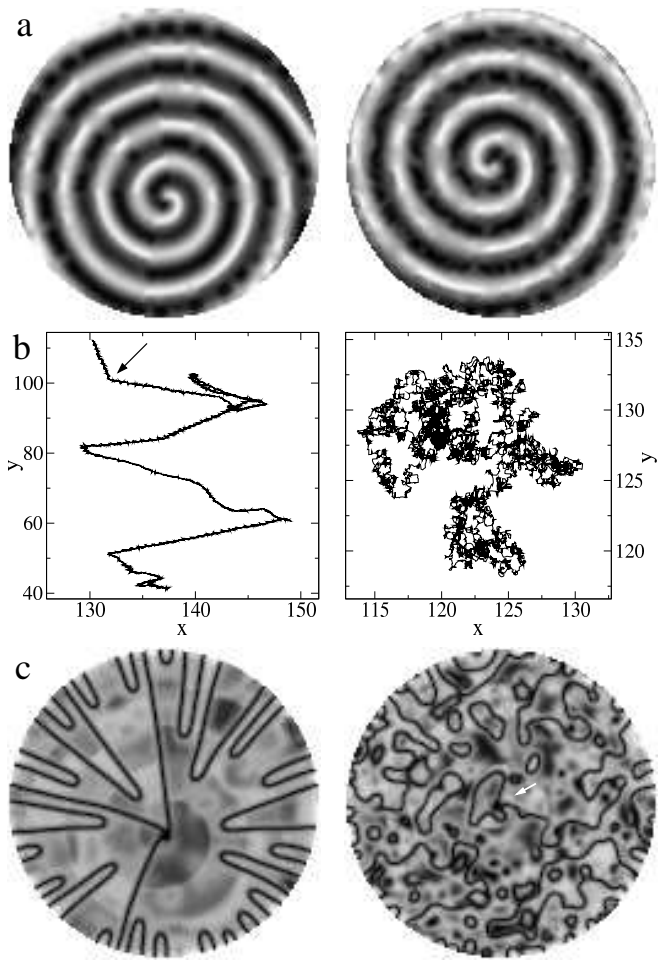}
 \caption{\label{traj_flight} Dynamics in the
 turbulent regime close to onset ($C=4.6$, left) and far from
 onset ($C=5.8$, right) for $L=512$. (a): Snapshots of the $c_x$
 field. (b): Trajectories of the spiral core.
 The black arrow points to the position of the spiral core at
 the time when the snapshot of $\Delta c_z$  was taken for
 $C=4.6$ (shown in c).
 (c): Snapshot of $\Delta c_z$ visualizing the line defects.
 Note that in the configurations shown three P1
 SDL are attached to the core.
 The white arrow points to the position of the spiral core.
 }
 \end{figure}

Next, we consider the turbulent regimes where P1 SDL are
spontaneously created ($4.557 < C \leq 6.0$). The continuous
creation and annihilation of P1 SDL and their dynamics strongly
influences the motion of the core. More complicated configurations
arise, which involve more than one P1 SDL attached to the core,
and the motion is no longer characterized by a simple angle
$\alpha$. Complex connection and reconnection events between P1
SDL as well as between SDL and the core continuously occur.
Close to the onset of this regime,
the trajectory often changes its direction abruptly, even
though long periods of (apparent) ballistic motion still persist.
When a single P1 SDL is attached to the core, the motion
is ballistic with $v\geq0.0017$ and $\alpha$ close to
$90^\circ$. Yet, most of the time, the spiral core has three
P1 SDL attached to it (Fig.~\ref{traj_flight}c) and
short-lived configurations with a
higher odd number of SDL exist as well. Even though a
configuration with three SDL is unstable for $C<4.557$, it has a
long lifetime and the combination of this persistence, the
continuous creation of P1 SDL, and the interaction with other
nearby SDL (Fig.~\ref{traj_flight}c), explains why this
configuration dominates for $4.557 < C \leq 6.0$. Interestingly,
the direction of motion can change without any apparent
variation in the configuration of the three attached P1 SDL. For
instance, during the change of direction highlighted by the arrow
in Fig.~\ref{traj_flight}b, the core configuration
shown in Fig.~\ref{traj_flight}c does not change significantly and
the three SDL remain attached to the core. This indicates that
part of the core motion is influenced by the dynamics of the
background field in the turbulent regime and/or that subtle
rearrangements occur deep inside the core.

 \begin{figure}
 \includegraphics[width=\columnwidth]{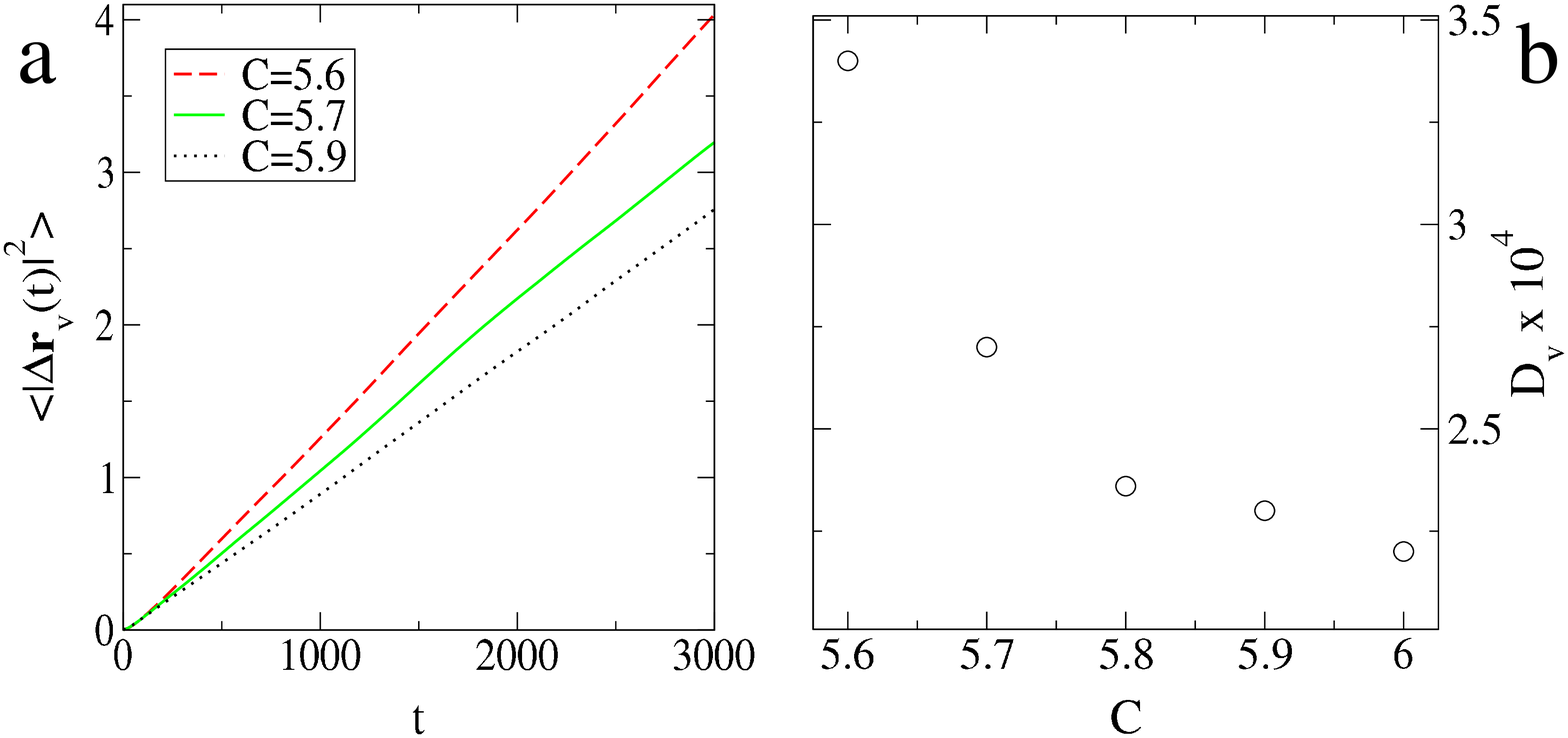}
 \caption{\label{diff} (color online) (a): Mean squared displacement
 $\langle|\Delta {\bf r}(t)|^2\rangle$ for different values of $C$ in the turbulent
 regime. (b): The core diffusion coefficient $D_v$ as a function of
 $C$.
 }
 \end{figure}

Deeper in the turbulent regime, the (apparent) ballistic flights
become shorter and shorter and the core trajectory resembles
Brownian motion (see Fig.~\ref{traj_flight}b). In this regime, SDL
are generated rapidly and homogeneously in the medium. This makes
it possible to characterize the core motion by a well-defined
diffusion constant $D_v$. Figure~\ref{diff}a shows the
mean-squared displacement $\langle|\Delta {\bf
r_v}(t)|^2\rangle=\langle|{\bf r_v}(t)-{\bf r_v}(0)|^2\rangle$ for
different values of $C$. Here, $\langle\cdots\rangle$ denotes a
time average over a long trajectory \cite{long} and ${\bf
r_v}(t)$ is the position of the core at time $t$. The dependence
of $\langle|\Delta {\bf r_v}(t)|^2\rangle$ on $t$ is clearly
linear as expected for Brownian motion. The diffusion constant of
the core motion determined from $\langle|\Delta {\bf
r_v}(t)|^2\rangle = 4 D_v t$ decreases with increasing $C$ (see
Fig.~\ref{diff}b), reflecting the fact that changes of direction
become more frequent while the mean velocity remains roughly
constant.

 \begin{figure}
 \includegraphics[width=\columnwidth]{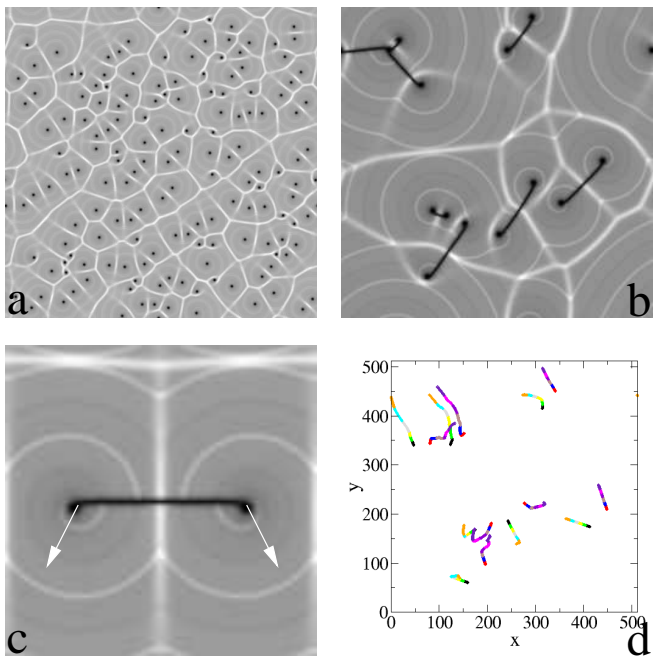}
 \caption{\label{multi} (color online)
 (a): Snapshot of $\Delta c_z$ showing a
 typical multi-spiral configuration in the P1 regime
 ($C=2.5$, $L=1024$).
  (b): same as (a) but in the P2 regime, where P1 SDL connect
 spiral cores ($C=3.5$, $L=512$)
 (c): same as (b) but for a single pair of spirals connected by a
 P1 SDL ($L=256$). The direction of the cores' motion is
 highlighted by the white arrows.
 (d): Trajectories of spiral cores leading to the configuration shown in
 (b). Every 3000 time units the color changes starting with red
 or black depending on the orientation of the spiral.
 }
 \end{figure}

We now turn to multispiral, spatially-disordered configurations
which occur spontaneously in any large-enough domain. For
oscillatory media described by the complex Ginzburg-Landau
equation, it was shown recently that spatially-disordered frozen
solutions do {\it not} exist \cite{brito03}. Even when the single
spiral solution is stable, the weak but non-trivial effective
interaction between spirals gives rise to ultra-slow core motion.
These results should apply to the R\"ossler medium in the simple
oscillatory P1 regime ($C\in[2.0, 3.03]$). A long run
performed at $C=2.5$ \cite{long} showed that after a transient
period a residual and barely detectable core motion subsists, with
typical velocities decreasing sharply with increasing spiral
domain size indicating a rapidly decaying interaction. This
observation, together with the wide variation of spiral domain
sizes (Fig.~\ref{multi}a), suggests that the P1 R\"ossler medium
is in the so-called ``vortex glass'' phase, and not the ``liquid''
phase where all spiral domains have roughly the same size. Thus,
core motion exists also in this regime without SDL, albeit with
velocities orders of magnitude smaller than those characteristic
of core motion due to complex oscillations and can thus be
neglected in the analysis of this latter case.

In a spatial domain with periodic boundary conditions there must
be an equal number of ``positive'' (clockwise) and ``negative''
(counter-clockwise) spirals. Under P2 local dynamics,
spirals have a tendency to be grouped in pairs (mostly of
opposite sign) connected by a P1 SDL (Fig.~\ref{multi}b).
Inevitably, the pattern rearranges since a connected pair drifts
apart as shown in Fig.~\ref{multi}c: a single spiral pair in a
square domain is connected by a
straight P1 SDL. The two cores move with speed $v$ at angles $\pm
\alpha$ relative to the SDL and drift away from each other with
speed $2v \cos(\pi-\alpha)$ which is positive as long as
$\left|\alpha\right|>90^\circ$. The sign of $\alpha$ is determined
by the sign of the spiral. The dependence of $v$ and $\alpha$ on
$C$ is indistinguishable from the single-spiral case
(Fig.~\ref{alpha}).

In multispiral disordered P2 regimes our results show that spiral
cores move nearly independently, stretching the P1 SDL to which
they are attached, until they meet another core leading to
reconnections and creation/annihilation of SDL. Some
``frustration'' arises during this process producing spiral cores
connected by more than one SDL. The rearrangements occur mainly
when two cores are close and the cascade of reconnection events
can span a distance of several wavelengths. In a P1 medium these
cores would annihilate. On the contrary, here, because $\alpha
> 90^\circ$, they repel each other after a new SDL connects them
and the annihilation of spiral pairs is prevented (see
Fig.~\ref{multi}d). Thus, multispiral configurations in P2 media
are characterized by a novel type of spatiotemporal chaos which
preserves the number of spiral cores through the complex dynamics
of the P1 SDL connecting them.

Simulations show that this new ``vortex liquid'' state exists
beyond the P2 regime, as long as $\alpha$ is significantly larger
than $90^\circ$. For $\alpha$ close to $90^\circ$, annihilation of
spiral pairs can occur, presumably due to the weaker repulsion
effect induced by the P1 SDL. Crossing the transition point to the
turbulent regime at $C=4.557$, the annihilation rate of spiral
pairs drastically increases. This is caused by the continuous
generation of new P1 SDL which leads to permanent reconnections
and effectively turns off the repulsion mechanism due to P1 SDL.

The new types of spiral core motion and turbulence described here
should be observable in distributed media with underlying complex
dynamics. Studies of the more realistic Willamowski-R\"ossler
reaction scheme \cite{wr} support this statement. We also expect
core motion to arise in excitable media supporting complex return
to the rest state \cite{excitable}. Experimentally,
Belousov-Zhabotinsky reaction-diffusion systems are especially
good candidates for studies of these phenomena since
complex-periodic and chaotic regimes with SDL have already been
observed in the laboratory \cite{park99}.

Research of RK supported in part by NSERC.


\begin{thebibliography}{19}
\expandafter\ifx\csname
natexlab\endcsname\relax\def\natexlab#1{#1}\fi
\expandafter\ifx\csname bibnamefont\endcsname\relax
  \def\bibnamefont#1{#1}\fi
\expandafter\ifx\csname bibfnamefont\endcsname\relax
  \def\bibfnamefont#1{#1}\fi
\expandafter\ifx\csname citenamefont\endcsname\relax
  \def\citenamefont#1{#1}\fi
\expandafter\ifx\csname url\endcsname\relax
  \def\url#1{\texttt{#1}}\fi
\expandafter\ifx\csname
urlprefix\endcsname\relax\def\urlprefix{URL }\fi
\providecommand{\bibinfo}[2]{#2}
\providecommand{\eprint}[2][]{\url{#2}}

\bibitem[{\citenamefont{Kapral and Showalter}(1995)}]{kapral}
\bibinfo{editor}{\bibfnamefont{R.}~\bibnamefont{Kapral}} \bibnamefont{and}
  \bibinfo{editor}{\bibfnamefont{K.}~\bibnamefont{Showalter}}, eds.,
  \emph{\bibinfo{title}{Chemical waves and patterns}}
  (\bibinfo{publisher}{Kluwer Academic Publishers},
  \bibinfo{address}{Dordrecht, Netherlands},
  \bibinfo{year}{1995});

\bibitem[{\citenamefont{Mikhailov}(1994)}]{mikhailov}
\bibinfo{author}{\bibfnamefont{A.}~\bibnamefont{Mikhailov}},
  \emph{\bibinfo{title}{Foundations of synergetics I: Distributed active
  systems}} (\bibinfo{publisher}{Springer Verlag}, \bibinfo{address}{New York},
  \bibinfo{year}{1994});
\bibinfo{author}{\bibfnamefont{D.}~\bibnamefont{Walgraef}},
  \emph{\bibinfo{title}{Spatio-temporal pattern formation}}
  (\bibinfo{publisher}{Springer Verlag}, \bibinfo{address}{New York},
  \bibinfo{year}{1997}).

\bibitem[{\citenamefont{Imbihl and Ertl}(1995)}]{imbihl95}
\bibinfo{author}{\bibfnamefont{R.}~\bibnamefont{Imbihl}} \bibnamefont{and}
  \bibinfo{author}{\bibfnamefont{G.}~\bibnamefont{Ertl}},
  \bibinfo{journal}{Chemical Reviews} \textbf{\bibinfo{volume}{95}},
  \bibinfo{pages}{697} (\bibinfo{year}{1995}).

\bibitem[{\citenamefont{Winfree}(1987)}]{winfree87}
\bibinfo{author}{\bibfnamefont{A.~T.} \bibnamefont{Winfree}},
  \emph{\bibinfo{title}{When time breaks down}} (\bibinfo{publisher}{Princeton
  University Press, Cambridge, NJ}, \bibinfo{year}{1987}).

\bibitem[{\citenamefont{Goldbeter}(1996)}]{goldbeter}
\bibinfo{author}{\bibfnamefont{A.}~\bibnamefont{Goldbeter}},
  \emph{\bibinfo{title}{Biological Oscillations and Cellular Rhythms}}
  (\bibinfo{publisher}{Cambridge University Press}, \bibinfo{year}{1996}).

\bibitem[{\citenamefont{Brusch et~al.}(2003)\citenamefont{Brusch, Torcini,
  and B{\"a}r}}]{brusch03}
\bibinfo{author}{\bibfnamefont{L.} \bibnamefont{Brusch}},
  \bibinfo{author}{\bibfnamefont{A.}~\bibnamefont{Torcini}},
  \bibnamefont{and} \bibinfo{author}{\bibfnamefont{M.} \bibnamefont{B{\"a}r}},
  \bibinfo{journal}{Phys. Rev. Lett.} \textbf{\bibinfo{volume}{91}},
  \bibinfo{pages}{108302} (\bibinfo{year}{2003}).

\bibitem[{\citenamefont{Gray et~al.}(1995)\citenamefont{Gray, Jalife, Panfilov,
 Baxter, Cabo, Davidenko, and Pertsov}}]{moveheart}
  \bibinfo{author}{\bibfnamefont{R.~A.} \bibnamefont{Gray}},
  \bibinfo{author}{\bibfnamefont{J.} \bibnamefont{Jalife}},
  \bibinfo{author}{\bibfnamefont{A.~V.} \bibnamefont{Panfilov}},
  \bibinfo{author}{\bibfnamefont{W.~T.} \bibnamefont{Baxter}},
  \bibinfo{author}{\bibfnamefont{C.} \bibnamefont{Cabo}},
  \bibinfo{author}{\bibfnamefont{J.~M.} \bibnamefont{Davidenko}},
  \bibnamefont{and} \bibinfo{author}{\bibfnamefont{A.~M.} \bibnamefont{Pertsov}},
  \bibinfo{journal}{Science} \textbf{\bibinfo{volume}{270}},
  \bibinfo{pages}{1222} (\bibinfo{year}{1995}).

\bibitem[{\citenamefont{Barkley}(1995)}]{barkley95}
\bibinfo{author}{\bibfnamefont{D.}~\bibnamefont{Barkley}}, in Ref. \cite{kapral}.

\bibitem[{\citenamefont{Aranson and Kramer}(2002)}]{aranson02}
\bibinfo{author}{\bibfnamefont{I.~S.} \bibnamefont{Aranson}} \bibnamefont{and}
  \bibinfo{author}{\bibfnamefont{L.}~\bibnamefont{Kramer}},
  \bibinfo{journal}{Rev. Mod. Phys.} \textbf{\bibinfo{volume}{74}},
  \bibinfo{pages}{99} (\bibinfo{year}{2002}).

\bibitem[{\citenamefont{Aranson et~al.}(1998)\citenamefont{Aranson, Chat{\'e},
  and Tang}}]{aranson98}
\bibinfo{author}{\bibfnamefont{I.~S.} \bibnamefont{Aranson}},
  \bibinfo{author}{\bibfnamefont{H.}~\bibnamefont{Chat{\'e}}},
  \bibnamefont{and} \bibinfo{author}{\bibfnamefont{L.-H.} \bibnamefont{Tang}},
  \bibinfo{journal}{Phys. Rev. Lett.} \textbf{\bibinfo{volume}{80}},
  \bibinfo{pages}{2646} (\bibinfo{year}{1998}).

\bibitem[{\citenamefont{Alonso and Sagu{\'e}s}(2001)}]{alonso01}
\bibinfo{author}{\bibfnamefont{S.}~\bibnamefont{Alonso}} \bibnamefont{and}
  \bibinfo{author}{\bibfnamefont{F.}~\bibnamefont{Sagu{\'e}s}},
  \bibinfo{journal}{Phys. Rev. E} \textbf{\bibinfo{volume}{63}},
  \bibinfo{pages}{046205} (\bibinfo{year}{2001}).

\bibitem[{\citenamefont{Scott}(1991)}]{scott}
\bibinfo{author}{\bibfnamefont{S.}~\bibnamefont{Scott}},
  \emph{\bibinfo{title}{Chemical chaos}} (\bibinfo{publisher}{Oxford University
  Press}, \bibinfo{address}{New York}, \bibinfo{year}{1991});
\bibinfo{author}{\bibfnamefont{J.-C.} \bibnamefont{Roux}} \bibnamefont{and}
  \bibinfo{author}{\bibfnamefont{H.~L.} \bibnamefont{Swinney}}, in
  \emph{\bibinfo{booktitle}{Nonlinear Phenomena in Chemical Dynamics}},
  \bibinfo{editor}{\bibfnamefont{C.}~\bibnamefont{Vidal}} \bibnamefont{and}
  \bibinfo{editor}{\bibfnamefont{A.}~\bibnamefont{Pacault}}, eds.,
  (\bibinfo{publisher}{Springer}, \bibinfo{address}{New York},
  \bibinfo{year}{1981});
\bibinfo{author}{\bibfnamefont{J.~L.} \bibnamefont{Hudson}} \bibnamefont{and}
  \bibinfo{author}{\bibfnamefont{J.~C.} \bibnamefont{Mankin}},
  \bibinfo{journal}{J. Chem. Phys.} \textbf{\bibinfo{volume}{74}},
  \bibinfo{pages}{6171} (\bibinfo{year}{1981}).

\bibitem[{\citenamefont{Goryachev et~al.}(2000)\citenamefont{Goryachev, Kapral,
  and Chat{\'e}}}]{goryachev00}
\bibinfo{author}{\bibfnamefont{A.}~\bibnamefont{Goryachev}},
  \bibinfo{author}{\bibfnamefont{R.}~\bibnamefont{Kapral}}, \bibnamefont{and}
  \bibinfo{author}{\bibfnamefont{H.}~\bibnamefont{Chat{\'e}}},
  \bibinfo{journal}{Int. J. Bifurcation Chaos Appl.
  Sci. Eng.} \textbf{\bibinfo{volume}{10}},
  \bibinfo{pages}{1537} (\bibinfo{year}{2000}).

\bibitem[{\citenamefont{Yonemama et~al.}(1995)\citenamefont{Yonemama, Fujii,
  and Maeda}}]{yoneyama95}
\bibinfo{author}{\bibfnamefont{M.}~\bibnamefont{Yonemama}},
  \bibinfo{author}{\bibfnamefont{A.}~\bibnamefont{Fujii}}, \bibnamefont{and}
  \bibinfo{author}{\bibfnamefont{S.}~\bibnamefont{Maeda}},
  \bibinfo{journal}{J. Am. Chem. Soc.}
  \textbf{\bibinfo{volume}{117}}, \bibinfo{pages}{8188} (\bibinfo{year}{1995}).

\bibitem[{\citenamefont{Park and Lee}(1999)}]{park99}
\bibinfo{author}{\bibfnamefont{J.-S.} \bibnamefont{Park}} \bibnamefont{and}
  \bibinfo{author}{\bibfnamefont{K.~J.} \bibnamefont{Lee}},
  \bibinfo{journal}{Phys. Rev. Lett.} \textbf{\bibinfo{volume}{83}},
  \bibinfo{pages}{5393} (\bibinfo{year}{1999});
\textbf{\bibinfo{volume}{88}},
  \bibinfo{pages}{224501} (\bibinfo{year}{2002}).

\bibitem{size}
Our results are independent of $L$ for $256 \le L \le 1024$ and of
the discretization $\Delta x$ for $0.5 \le \Delta x \le 1.0$.

\bibitem[{\citenamefont{R{\"o}ssler}(1976)}]{roessler76}
\bibinfo{author}{\bibfnamefont{O.}~\bibnamefont{R{\"o}ssler}},
  \bibinfo{journal}{Phys. Lett. A} \textbf{\bibinfo{volume}{57}},
  \bibinfo{pages}{397} (\bibinfo{year}{1976}).

\bibitem[{\citenamefont{Sandstede}(2001)}]{sandstede}
\bibinfo{author}{\bibfnamefont{B.}~\bibnamefont{Sandstede}},
    private communication, (\bibinfo{year}{2001}).

\bibitem{long}
Trajectory length was approximately $10^5$ spiral periods.


\bibitem[{\citenamefont{Brito et~al.}(2003)\citenamefont{Brito, Aranson, and
  Chat{\'e}}}]{brito03}
\bibinfo{author}{\bibfnamefont{C.}~\bibnamefont{Brito}},
  \bibinfo{author}{\bibfnamefont{I.~S.} \bibnamefont{Aranson}},
  \bibnamefont{and}
  \bibinfo{author}{\bibfnamefont{H.}~\bibnamefont{Chat{\'e}}},
  \bibinfo{journal}{Phys. Rev. Lett.} \textbf{\bibinfo{volume}{90}},
  \bibinfo{pages}{068301} (\bibinfo{year}{2003}).


\bibitem[{\citenamefont{Willamowski and R{\"o}ssler}(1980)}]{wr}
\bibinfo{author}{\bibfnamefont{K.~D.}~\bibnamefont{Willamowski}}
  \bibnamefont{and}
  \bibinfo{author}{\bibfnamefont{O.~E.}~\bibnamefont{R{\"o}ssler}},
  \bibinfo{journal}{Z. Naturforsch.} \textbf{\bibinfo{volume}{35}},
  \bibinfo{pages}{317} (\bibinfo{year}{1980}).

\bibitem[{\citenamefont{Goryachev and Kapral}(1999)}]{excitable}
\bibinfo{author}{\bibfnamefont{A.}~\bibnamefont{Goryachev}}, \bibnamefont{and}
  \bibinfo{author}{\bibfnamefont{R.}~\bibnamefont{Kapral}},
  \bibinfo{journal}{Int. J. Bifurcation Chaos Appl.
  Sci. Eng.} \textbf{\bibinfo{volume}{9}},
  \bibinfo{pages}{2243} (\bibinfo{year}{1999}).

\end{thebibliography}
\end{document}